# Probing of Structural Phase Transitions in Barium Titanate Modified Sodium Niobate using Raman Scattering

Mrinal Jauhari[1,2], S. K. Mishra[1], R. Mittal[1,2] and S. L. Chaplot[1,2]

[1]*Solid State Physics Division, Bhabha Atomic Research Centre, Trombay, Mumbai 400 085, India*
[2]*Homi Bhabha National Institute, Anushaktinagar, Mumbai 400 094, India*

**Abstract**

Raman Spectroscopic measurements are carried out to investigate the structural phase transitions as a function of composition in modified sodium niobate [(1-x) $NaNbO_3$-$xBaTiO_3$:NNBTx] for $x$=0.0 to 0.15 at room temperature. The characteristic antiferroelectric modes at around 93.4 and 123.6 $cm^{-1}$ alongwith a mode at 155.5 $cm^{-1}$ were found to disappear across the structural phase transition from antiferroelectric orthorhombic phase (*Pbcm*) to ferroelectric orthorhombic phase (*Pmc2$_1$*) phase for $x$>0.02. The redistribution of intensities and positions of the Raman lines in bending (150-350 $cm^{-1}$) and stretching modes (>550 $cm^{-1}$) on increasing the concentration $x$>0.05 also confirms the occurrence of another phase transition from ferroelectric orthorhombic phase (*Pmc2$_1$*) to another ferroelectric orthorhombic phase (*Amm2*) phase across x~0.10. The phase transitions as observed from Raman measurements are consistent with previous x-ray diffraction study.

**Introduction**

Development of lead-free and eco-friendly piezoelectric materials is a necessity for the next generation piezoelectric sensor and actuators[1,2]. The physical properties and performance of these materials are governed by their structural phases at desired temperatures. The correct knowledge of structural phase transitions helps us to develop a suitable material for a particular application and for the enhancement of the device performance. In this regard, sodium niobate based lead-free piezoelectrics are the one of the most promising materials[1]. Sodium niobate is also the most studied material for understanding the structural phase transitions in Perovskite $ABO_3$ type compounds. It shows several phase transitions from high-temperature non-polar paraelectric cubic (**Pm$\bar{3}$m**) to low-temperature polar ferroelectric rhombohedral (**R3c**) phase. Above 913 K it has a paraelectric cubic phase (Pm$\bar{3}$m). On lowering the temperature, it undergoes transition to a series of antiferrodistortive phases in this order: tetragonal (T2) P4/mbm, orthorhombic (T1) Cmcm, orthorhombic (S) Pbnm, orthorhombic (R) Pbnm, orthorhombic (P) Pbcm phases, and a rhombohedral R3c phase.[3,4] Pressure-induced phase transition from room-temperature antiferroelectric orthorhombic (**Pbcm**) phase to paraelectric orthorhombic (**Pbnm**) phase is also reported in literature.[3-7] These structural phase transitions involve phonon mode instabilities linked with zone centre and zone boundary phonon modes with respect to the cubic phase.[3,4,6-11] Zone-centre mode instability is associated with off centring of cations with respect to oxygen octahedra resulting information of dipoles and promotes ferroelectricity in the material. However, the zone-boundary instabilities are associated with rotations of the oxygen octahedra.

The off centring of cations and rotations and/or tilting of oxygen octahedra can be tailored by substitution of atoms at the A and B sites of the $ABO_3$ structure. For example, on doping of Li, K etc[12,13] at A site, the room temperature antiferroelectric phase of $NaNbO_3$ get suppressed, and the doping promotes polar distortions, i.e., relatively large B site cationic displacement inside the oxygen octahedra. The incorporation of relatively large A site atom increases the overall volume of the unit cell and reduces the amount of rotations and/or tilting of oxygen octahedra. Moreover, it is also well-known fact that the solid solutions of antiferroelectric and ferroelectric compounds show competing behaviour and induce large spontaneous polarization such as in $(1-x)PbZrO_3-xPbTiO_3$ (PZT), $K_{(1-x)}Na_xNbO_3$ (KNN), $Li_{(1-x)}Na_xNbO_3$ (LNN) etc**.** By considering all these facts, we have intentionally prepared the solid solution of $NaNbO_3$ with $BaTiO_3$, which shows ferroelectricity at room temperature. This combination of antiferroelectric ($NaNbO_3$) and ferroelectric ($BaTiO_3$) compounds is similar to $(1-x)PbZrO_3-xPbTiO_3$ (PZT) and we are expecting similar high electromechanical response. Here the $Ba^{2+}$ is larger in size as compared to $Na^+$, and $Ti^{4+}$ is little bit smaller than the $Nb^{5+}$, which induces the distortion at A and B sites in the $NaNbO_3$ matrix and doping of $BaTiO_3$ enhances the electrical response of these solid solutions $(1-x)NaNbO_3$-$xBaTiO_3$:(NNBT$x$).

The dielectric, ferroelectric and piezoelectric properties of NNBT$x$ [$(1-x)NaNbO_3$-$xBaTiO_3$] ceramics have been investigated by various workers[6,14-16]. They have shown that on increasing the doping of $BaTiO_3$ in $NaNbO_3$, there is a significant increment in the dielectric and piezoelectric response of material. Recently, using a combination of dielectric and powder x-ray and neutron diffraction studies[6,17], we have shown that on increment of $BaTiO_3$ concentration in $NaNbO_3$ matrix, the parent antiferroelectric orthorhombic phase with space group ***Pbcm*** (**AFE**) is destabilized and it transforms to the ferroelectric orthorhombic phase with space group ***Pmc2$_1$*** (**FE1**) at x>0.02. On further increase in the doping at x~0.10, the sample transforms to another orthorhombic phase with space group ***Amm2*** (**FE2**)[6,17]. We have also observed the increment in the dielectric constant with composition.

For a proper understanding of the properties of the materials, the structural information as obtained from diffraction techniques should be supplemented and supported by other techniques. The structural phase transitions in these compounds are governed by softening of various phonon modes, which can be probed via various experimental methods including Raman spectroscopy, infrared spectroscopy and inelastic neutron scattering. Raman spectroscopy is sensitive to local structure changes and is a suitable tool to investigate the phase transitions in these materials.[16,18-25] With this aim in mind, we performed Raman spectroscopy measurements to provide a deep insight of the structural phase transitions and the role of phonons on changing the composition.

**Experimental**:

Polycrystalline $(1-x)$ $NaNbO_3$–$xBaTiO_3$: NNBT$x$ samples with composition $x$=0.0, 0.005, 0.01, 0.02, 0.03, 0.05, 0.07, 0.1, 0.12, 0.15 and 0.20 are prepared by solid–state thermochemical reaction method and details are given in Supplementary Materials. The Raman spectra were recorded in a micro Raman spectrograph LabRAM HR 800 evolution with a 50X LWD objective. The excitation source used was 532 nm solid state laser of power ~ 25mW. The Raman instrument has spectral resolution < 1.5 cm$^{-1}$. The spot size of the beam is of 3-4 μm. The ultra-low frequency cut-off filter is used to record the spectra from 10 cm$^{-1}$. Usually Raman line shapes of weakly anharmonic phonons are fitted with Lorentzian functions. However, soft modes may be highly anharmonic, which may be fitted with a damped harmonic oscillator model. We have fitted the Raman modes with damped harmonic oscillator (DHO) model[26]. The central peak is fitted with a Lorentz function. The combined function used for the entire Raman spectra with the central peak is given below with constant baseline.

$$I(\omega) = \frac{A_{CP}(\Gamma_{CP}/2)}{\pi[\omega^2 + (\Gamma_{CP}/2)^2]} + [n(\omega) + 1] \sum_{k=1}^{n} \frac{A_k \Gamma_k \omega_k^2 \omega}{[(\omega^2 - \omega_k^2)^2 + (\omega \Gamma_k)^2]}$$

Where, $n(\omega)$ is the Bose-Einstein population factor.

$$n(\omega) = \frac{1}{[e^{hc/\lambda kT} - 1]}$$

A and $\Gamma_{CP}$ are the proportional constant and the width of central peak, respectively. $A_k$, $\omega_k$, and $\Gamma_k$ are the amplitude, peak position, and damping constant (linewidth), respectively, of the k$^{th}$ Raman mode.

## Results:

### A. Group theoretical analysis for Raman active modes:

Recently, we have studied dielectric and structural properties of NNBTx using a combination of impedance spectroscopy and x-ray diffraction techniques[6]. Based on the appearance and/or disappearance of the superlattice reflections in the x-ray diffraction data, we observed that on increasing concentration of $BaTiO_3$ in $NaNbO_3$ matrix, there are drastic changes in the intensities of main perovskite peaks and superlattice reflections on increasing the dopant concentration (see Fig. S1 (Supporting Information). The X- ray diffraction patterns for upto x<0.02 could be indexed using the AFE (**Pbcm**) phase. On increasing the concentration *x*≥ 0.02, we found that the superlattice reflection around 36.2˚ could not be indexed using the same AFE phase model and crystal structure transforms from antiferroelectric orthorhombic AFE(**Pbcm**) phase to ferroelectric orthorhombic FE1 (**Pmc2₁**)phase. Further, increase in the dopant concentration leads to suppression of another superlattice reflection S2 at 2θ~ 38.5˚ and diffraction pattern could be indexed[6] using ferroelectric orthorhombic FE2 (**Amm2**) phase for *x*≥0.10. The peaks marked with asterisk correspond to $Ba_3Nb_4Ti_4O_{21}$ and the results of two phase refinement suggest that the content of the undesired $Ba_3Nb_4Ti_4O_{21}$ phase is about ~1% and it increases upto 3% on increasing $BaTiO_3$ content to x=0.20.

Details of the normal mode analysis performed using free access site Bilbao Crystallographic Server[27] for these three phases are given below:

1) **Orthorhombic AFE (for composition $0 \leq x < 0.03$):** At ambient condition, pure $NaNbO_3$ has AFE orthorhombic (**Pbcm**) symmetry with unit cell dimensions $\sqrt{2}a_P \times \sqrt{2}a_P \times 4a_P$ (with respect to cubic perovskite with lattice parameter $a_P$). It has eight numbers of formula units (total 40 atoms) and total 120 possible normal modes. The asymmetric unit contain two types of sodium atoms (Na), one type of niobium atom (Nb) and four types of oxygen atoms. Table S1(Supporting Information) contains the details of the group theoretical analysis for the AFE orthorhombic phase.

2) **Orthorhombic FE1 (for composition $0.02 < x < 0.07$):** The structure of the NNBT$x$($x=0.02 < x < 0.07$) at room temperature was found to be orthorhombic (**Pmc2$_1$**) symmetry with unit cell dimensions $2a_P \times \sqrt{2}a_P \times \sqrt{2}a_P$ and four numbers of formula units. i.e. total 20 atoms. There should be total 60 possible normal modes. The group theoretical analysis gives 57 Raman active modes represented by $\Gamma_{Raman}=16A_1+13A_2+12B_1+16B_2$. These Raman active modes are simultaneously IR active due to non-centro symmetry of the crystal structure except $A_2$ mode. Table S2 (Supporting Information) contains the details of the group theoretical analysis for the FE1 orthorhombic phase.

3) **Orthorhombic FE2 (for composition $0.10 \leq x \leq 0.15$):** The structure of the NNBT10 at room temperature was found to be orthorhombic (**Amm2**) symmetry with cell dimensions $a_P \times 2a_P \times 2a_P$ and four number of formula units, i.e., total 20 atoms. In primitive unit cell there are 10 atoms. There should be total 30 possible normal modes. The group theoretical analysis gives 27 Raman active modes represented by $\Gamma_{Raman}=9A_1+2A_2+7B_1+9B_2$. Table S3(Supporting Information) contains the details of the group theoretical analysis for the FE2 orthorhombic phase.

**Raman Spectroscopy:**

$NaNbO_3$ belongs to the $ABO_3$ type perovskite family and at highest temperature it has paraelectric cubic phase. This cubic phase unit cell contains six oxygen atoms, which reside at the face centred positions and form oxygen octahedron. Inside the void of the octahedron there is a metal cation B (in present case $Nb^{5+}$). Eight A type ($Na^+$) cations occupy the corners of the unit cell. This cubic (**Pm$\bar{3}$m**) phase is characterized by second order Raman scattering. At lower temperatures, the structure gets slightly distorted and the effect of these distortions changes the symmetry and number of formula units in unit cell in daughter phase. As a result, appearance or disappearance of Raman modes in the Raman spectra are observed and can be treated asa signature for structural phase transitions. We observed clear appearance and disappearance of first-order Raman modes in the Raman spectra of NNBT$x$ as a function of composition and details are discussed below.

Figure 1 shows the evolution of Raman spectra collected at ambient conditions as a function of composition in the range 10to 1000 cm$^{-1}$. Raman spectrum of NaNbO$_3$is similar as reported by various authors[18-20,28-31]. The spectra could be decomposed into two main regions involve vibrations of the Na$^+$ against the NbO$_6$ octahedra and the isolated vibrations of NbO$_6$ octahedra, namely, the external and internal modes. Under the consideration of external modes, the vibrational bands below 100 cm$^{-1}$are the translational modes (**Tr**),which belong to the motion of the Na$^+$ ions against the NbO$_6$ octahedra. The bands at 93.4, 123.6, 143.2 and 155.3 cm$^{-1}$ are the liberational modes (**L**) of NbO$_6$ octahedra. These are the roto-vibrational modes of the NbO$_6$ octahedra. The internal modes appear at relatively higher wave numbers and these comprise the bending of bond angles (O-Nb-O) and the stretching motion of the bonds (Nb-O) within the octahedron. Following the literature for the NbO$_6$ octahedron in the perovskite lattice structures[32], the vibrations of isolated equilateral octahedron consists of $A_{1g}(\upsilon_1)+E_g(\upsilon_2)+2F_{1u}(\upsilon_3, \upsilon_4)+F_{2g}(\upsilon_5)+F_{2u}(\upsilon_6)$ normal modes. The $\upsilon_1,\upsilon_2$, and $\upsilon_3$modes belong to the stretching motion and $\upsilon_4,\upsilon_5$, and $\upsilon_6$ arethe bending modes of a perfect octahedron[19,29]. Out of these six modes only $\upsilon_1,\upsilon_2$, and $\upsilon_5$ are Raman active and $\upsilon_3$ and $\upsilon_4$ are IR active. The triply-degenerate mode $\upsilon_6$ is a silent mode. However, depending on the site symmetry these modes get split and non-Raman active modes may become Raman active. For the spectrum of pure NaNbO$_3$,the bands located in the range 160 to 190, 190 to 300, 370 to 440 and around 560, 600, 670 cm$^{-1}$are assigned in literature as $\upsilon_6, \upsilon_5, \upsilon_4$ [bending modes (**B**)] and $\upsilon_2,\upsilon_1$, and $\upsilon_3$ [stretching modes (**S**)] respectively. Further, on increasing the dopant concentration, the entire Raman spectrum shifts towards the lower wavenumber side and dramatic changes are seen in the relative intensities of the librational and bending modes of the Raman profiles (shown in right hand side zoomed part of figure 1).The Raman spectra for the composition x<0.05 looks similar because both **AFE** (**Pbcm**) and **FE1** (**Pmc2$_1$**) phases have similar number of Raman active modes (60 and 57 Raman active mode). However, for the composition x≥0.5, the spectrum becomes diffusive as a result of doping and may be associated with increase of disorder in the crystal.

As said earlier, NNBT$x$ undergoes structural phase transitions around $x$>0.02 and 0.10. Therefore, we have taken $x$=0.0, 0.03 and 0.10 as a representative of these three phases namely: **AFE, FE1** and **FE2** phases, respectively. To quantify the Raman mode intensities and peak positions, we have fitted the whole Raman spectra using damped harmonic oscillator (DHO) model. Figure 2 (a) shows the fitting of the Raman spectra obtained for NaNbO$_3$ at ambient conditions. Total 22 peaks were used to get good fitting of the observed data. However, we estimated total 60 Raman active modes using group theoretical calculations. The fitted Raman bands showed good agreement with the previous reports on pure bulk sodium niobate. Figure 2 (b), (c) and (d) are close views of the translational (**Tr**), librational (**L**), bending (**B**) and stretching (**S**) modes of the same profile. Figure 2 (e) and (f) show the fitting of the observed Raman data of these compositions, respectively.

Normal mode analysis suggests 57 Raman active modes for the **FE1** (**P**$mc2_1$) phase, but 17 peaks gave a good fitting of the Raman spectra observed for NNBT03. The Raman spectrum looks similar as reported by Sheratori *et. al.*[29] For **FE2** (**A**$mm2$) phase, factor group analysis suggested a total of 27 Raman active modes, but only 12 peaks were sufficient for fitting of the observed broad Raman spectra of NNBT10. The number of calculated and fitted Raman modes may be different due to less resolved and broad features of the observed spectra. Similar broadening in the Raman spectra was also observed in PMN-PT solid solution and explained in terms of coexistence of polar nanoregions and chemically ordered clusters[25]. They show that presences of the polar nanoregions (PNR) are due to the ions at the B site: Nb, Mg and Ti being displaced from their high symmetry positions, give rise to local electric dipoles. The presence of chemically ordered clusters additionally complicates understanding of the Raman spectra. As the Raman cross section depends mainly on the bond polarizability, i.e., the number of electron involved in the bond, it is highly possible that the Raman spectra will not reflect the signature of all regions. In the case of PMN-PT, the following sequences of the Raman mode intensities are expected: Nb-O > Ti-O >> Mg-O. As a consequence, the contribution of Mg-rich regions is too weak to be observed, as far as the bending and stretching modes are considered. Consequently, the unambiguous Raman mode assignment is rather delicate.

Positions of the peaks for NNBT$x$ ($x$=0.0, 0.03 and 0.10) are compared with the previously reported values for bulk sodium niobate given in Table 1. The Raman profiles of the samples with x = 0.02, 0.07 and 0.09 indicate some minor features which suggest mix phases that are consistent with the x-ray diffraction results discussed below. We have analysed all the spectra and the variation of the peak positions as obtained through the fittings are shown in figure 2.

**Discussion:**

In pure NaNbO$_3$, the two main peaks due to translation of Na$^+$ ion against the NbO$_6^-$ octahedra clearly indicate that there are two types of interactions between these ions due to two distinct types of Na$^+$ ions (sitting at different sites). The separation between these two peaks decreases and both shift towards lower wave number side continuously on increasing the doping concentration. The broadening and merging of these peaks could be related to the presence of more types of interactions in the crystal due to the incorporation of different type of ions (Na$^+$ and Ba$^{+2}$) with different sizes. Also, the different type of atoms will show different types of bonding with oxygen atoms. The overall translational modes get broadened beyond our detection range on increase of dopant concentration $x>0.05$. The most prominent changes could be seen in the librational modes around wave number 93.4 cm$^{-1}$ and 155.5 cm$^{-1}$ (figure 1). These modes disappear at $x\sim0.03$, and all the librational modes get broadened at a higher concentration of dopant for $x>0.05$. In previous phonon dynamics study on NaNbO$_3$, Mishra *et. al.*[9] proposed that the Raman modes at around 93 cm$^{-1}$ and 123 cm$^{-1}$ belong to A$_{1g}$ modes and they appear due to folding of the corresponding specific zone centre points[9]. They identified that the mode at 93cm$^{-1}$ appears due to folding of T point in the Brillouin zone of the cubic phase and involves significant motion of Na, Nb, and O atoms located at 4d (¼+u, ¾+v, ¼), 8e (¼+u, ¼+v, ⅛+w), and (½+u, 0+v, ⅛+w) sites, respectively. This T point mode along with R$_{25}$ and M and Δ point mode instabilities is responsible for antiferroelectricity in NaNbO$_3$. In our investigation we observe that the modes around 93.4, 123.6 and 155.5 cm$^{-1}$ disappear when the structure transforms from orthorhombic **AFE** (**P*bcm***) phase to orthorhombic **FE1** (**P*mc2$_1$***) phase at $x\sim0.03$ and can be treated as a signature of the structural phase transition in this compound.

Bending modes υ$_6$, υ$_5$, and υ$_4$ of the NbO$_6$ octahedra are very much sensitive to the changes in the lattice parameters of the unit cell. We observed that the continuous expansion in the volume of the unit cell leads to increase in the average Nb-O bond length and corresponding decrease in the force constants. Due to this effect overall bending modes are shifted towards the lower wave number side with anomalous broadening with increasing the concentration of BaTiO$_3$ in NaNbO$_3$ matrix. The bending mode υ$_6$ arises due to the distortion of the perfect oxygen octahedra. In the case of perfect octahedra, we do not get the intensity of the υ$_6$ mode[19]. The presence of this band near 185 cm$^{-1}$ is clearly visible for pure NaNbO$_3$, which indicates that the NbO$_6^-$ octahedra are slightly distorted. However, the intensity of this band decreases continuously with increased broadening. This could be the result of Ti$^{+4}$ ion substitution with Nb$^{+5}$ and doping of the relatively large ion Ba$^{+2}$ at A site which increases the overall volume of the perovskite cell and making the octahedra less distorted. The internal bending modes $\upsilon_5$ become broad and converge with the $\upsilon_4$ at $x>0.05$. The continuous change in the relative intensities and broadening of these bending modes is observed.

Figure 3shows the variation of the peak positions and damping constant for the NNBTx solid solution (x=0.0 ≤x ≤0.15). Shift in peak position and damping constant of the main bending modes near about 277 and 255 cm$^{-1}$is remarkable (figure 3 (d) and 3 (e) respectively) and show anomalous variation around $x$=0.02 which indicates the changes in the microscopic environment of the $NbO_6^-$ octahedra due to the changes in the Nb-O bond lengths with cell expansion. The higher wave number modes$\upsilon_1$, $\upsilon_2$ and $\upsilon_3$ (which are due to stretching motion of internal octahedra) get broadened on increasing the dopant concentration. Also, we got only a very feeble signature of $\upsilon_3$ mode which indicates significant reduction in the distortions of $NbO_6$ octahedra for $x$ >0.05.Variation of the peak positions and damping constant of stretching modes present around 601 cm$^{-1}$ and 560 cm$^{-1}$are plotted in the Figures 3 (b) and (c) respectively. The peak position shows anomalous behaviour at $x$>0.02 and 0.10 with change in slope. All these modes are broad because of the disorder in the system at both A and B sites for higher concentrations.

To relate the results of Raman spectroscopic measurements with the x-ray diffraction study, we have obtained the structural parameters using Rietveld refinement of the x-ray diffraction data of these compositions as shown in Figure 4. We have plotted the cell parameters of these phases in terms of the cell parameters of the equivalent elementary perovskite (i.e., pseudo-cubic unit cell).We notice some phase coexistence for x= 0.02 and in the range 0.07≤x≤0.10, whichmay be due to first order nature of the phase transitions. In Figure4, we have plotted the structural parameters corresponding to the majority phases only.The variation of pseudo-cubic unit cell parameters clearly shows the monotonic increase in the lattice parameters and volume and jumps at phase transition boundaries (Figure 4(a)). Figure 4(b) shows the variation of volume of oxygen octahedra in the **AFE** (**P***bcm*) and **FE1** (**P***mc2$_1$*) phase. There is only one type of Nb atom and hence only one type of octahedra. However, in case of **FE2** (**A***mm2*), there are two types of Nb atoms. So, we get two types of octahedra present in the structure and all the octahedra dipoles are aligned in the same direction (i.e. Nb displacement is in the same direction). The volume of octahedra shows small change around $x$~0.02 when structure transforms from **AFE** to **FE1** phase and there is a bifurcation with abrupt change at around $x$~0.10, which corresponds to **FE2** phase. The axial Nb-O bond lengths of oxygen octahedra first increases slowly up to $x$~0.07 and then there is a huge jump and bifurcation due to the presence of two types of Nb atoms present in orthorhombic **FE2** (**A***mm2*) symmetry (Fig. 4 (c)). This increase in bond lengths is consistent with the low wave number shift of the bending modes obtained in Raman profile. The plot of variation of axial and planar Nb-O-Nb bond angles with BaTiO$_3$ content ($x$) (Figure 4(d)) suggests that first these angles changes little bit during the transition from **AFE** (**P$bcm$**) to **FE1** (**P$mc2_1$**) phase then there is almost linear increment in the axial bond angle and it becomes 180 degrees around $x$~0.10 which suggest that there is a suppression of one octahedra tilt. The planar Nb-O-Nb bond angle also increases discontinuously at $x$~0.10 which is related to the decrease in distortion in the NbO$_6$ octahedra on increasing the BaTiO$_3$ content in the NNBTx solid solution.

**Conclusions:**

In summary, we observed a distinct change in Raman profile across the composition $x$= 0.0, 0.03 and 0.10. The characteristic antiferroelectric mode at 93.4,123.6 alongwith a mode at 155.5 cm$^{-1}$ present in NaNbO$_3$ in Raman spectra, disappears on doping. These suggest the suppression of the antiferroelectric phase and confirms a structural phase transition across x>0.02. Further, on increasing the concentration of dopant above $x$>0.05, the Raman spectra become diffusive. We observed a distinct change in terms of redistribution of intensities and positions of Raman lines in bending (in range of 100-300 cm$^{-1}$) and stretching modes (>550 cm$^{-1}$). Shifting of Raman spectra towards lower wave number and change in the intensities of $\upsilon_6$ mode (which reflects a measure of the distortion in the NbO$_6$ octahedra) with increment of the dopant could be explained in terms of bond lengths and Nb-O-Nb bond angles. These results from Raman scattering are consistent with previous powder x-ray diffraction studies, which showed drastic changes in the intensities of main perovskite peaks and superlattice reflections on increasing concentration of BaTiO$_3$ in NaNbO$_3$ matrix.

**Acknowledgement:**

S. L. Chaplot would like to thank the Department of Atomic Energy, Government of India, for the award of Raja Ramanna Fellowship. The authors would like to thank Dr. Rekha Rao and Mr. Swayam Kesari, Raman spectroscopy facility, Solid State Physics Division, Bhabha Atomic Research Centre, Mumbai, India for measurement of Raman spectra.

**Table 1.** Comparison of the Raman band positions (cm$^{-1}$) observed in the present study for NNBT$x$ with the reported data of the NaNbO$_3$ in previous studies. The assignment of the modes has been made following the literature for vibrations of isolated equilateral NbO$_6$ octahedra in the perovskite lattice structures[18,32].

| NaNbO$_3$[19] AFE (Pbcm) Phase | Li$_{0.12}$Na$_{0.88}$NbO$_3$[13] FE (Pmc2$_1$) Phase | Present Study | | | |
|---|---|---|---|---|---|
| | | NaNbO$_3$ | NNBT03 | NNBT10 | Assignment |
| 60.2 | | 57.4 61.9 | 62.6 | 68.8 | Na$^+$ |
| 73.0 | 81 | 74.9 | 74.4 | ---- | |
| 89.5 | | 93.4 | ---- | ---- | NbO$_6^-$ libration |
| 115.1 | | ---- | ---- | ---- | |
| 121.0 | 120.1 | 123.6 | 116.5 | 101.5 | |
| 136.9 | | ---- | 137.2 | ---- | |
| 142.1 | | 143.3 | ---- | 143.1 | |
| | 149.1 | | | | |
| 153.4 | | 155.5 | ---- | | |
| 175.1 | | 177.9 | 173.7 | 166.8 | υ$_6$ (F$_{2u}$) |
| 183.2 | | 185.7 | ---- | ---- | |
| | 194 | | | | |
| 201.0 | | 201.6 | 195.9 | ---- | υ$_5$ (F$_{2g}$) |
| 218.8 | 226 | 220.4 | 222.8 | 211.4 | |
| 247.6 | | 256.2 | 249.4 | 234.4 | |
| 255.1 | 254 | 277.5 | 279.2 | 275.2 | |
| 276.1 | 285.4 | 296.0 | 308.8 | ---- | |
| 295.9 | | 324.0 | ---- | ---- | |
| 378.5 | | 380.7 | 359.3 | 335.3 | υ$_4$ (F$_{1u}$) |
| 435.3 | 436.7 | 435.4 | 435.8 | 437.9 | |
| 557.2 | 564.9 | 557.8 | 568.6 | 572.9 | υ$_2$ (E$_g$) |
| 602.6 | 618.4 | 602.4 | 611.2 | 615.0 | υ$_1$ (A$_{1g}$) |
| 671.5 | | 671.6 769.6 | 669.2 780.5 | 737.0 | υ$_3$ (F$_{1u}$) |
| 867.3 | 873 | 870.8 | 869.9 | 865.2 | υ$_5$ + υ$_1$ |

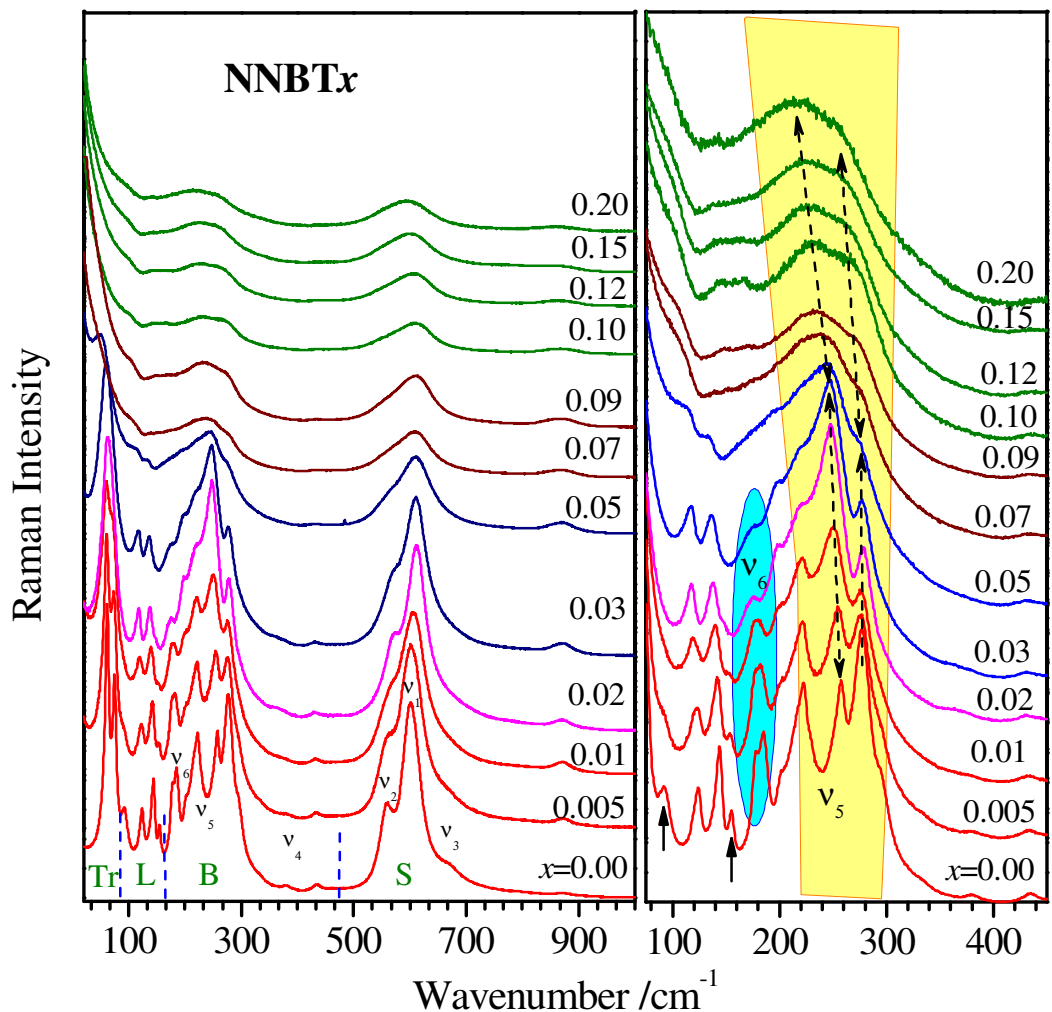

**Figure 1** (Left panel) Evolution of the Raman spectra collected in ambient conditions for various compositions of NNBTx (x=0.0 ≤x ≤0.20). The zoom view of the spectra up to 450 cm$^{-1}$ is shown in the right panel. For clarity, the intensity of each spectrum in the right panel is normalized to unity from 75 to 450 cm$^{-1}$. The Raman spectra may be grouped as translational (Tr), librational (L), bending (B) and stretching (S) modes. The region that shows maximum change in Raman spectra is highlighted by yellow and blue colours. The librational modes around wave number 93.4 and 155.3 cm$^{-1}$ are marked with arrows.

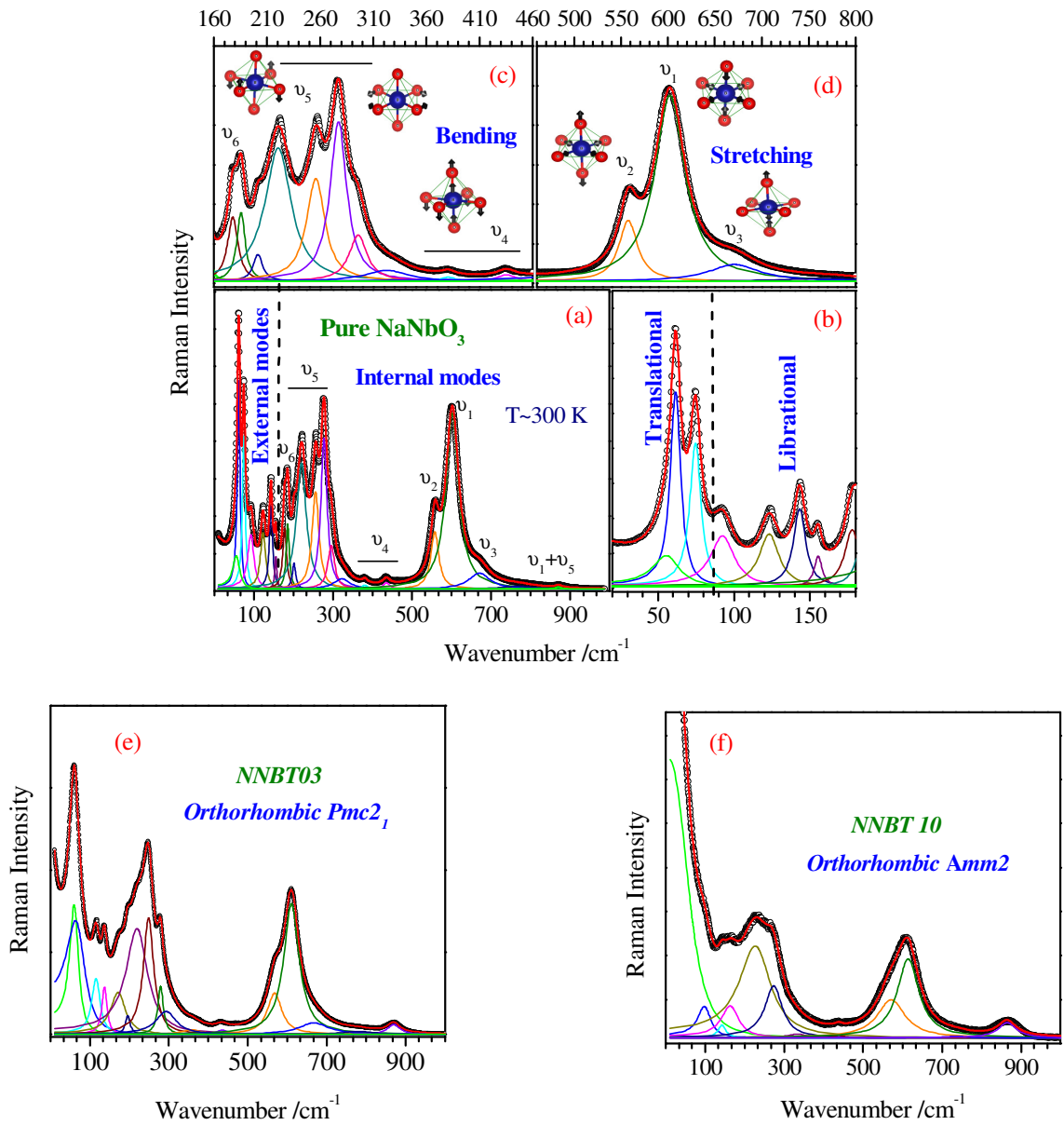

**Figure 2** Fitting of Raman spectroscopy data of pure (a) $NaNbO_3$, (e) NNBT03 and (f) NNBT10. Open circles (black) and solid line (red) represent the observed and fitted profile of Raman spectra. Figure 3 (b) (c) and (d) shows the expanded view of fitted Raman spectra of $NaNbO_3$ in different region.

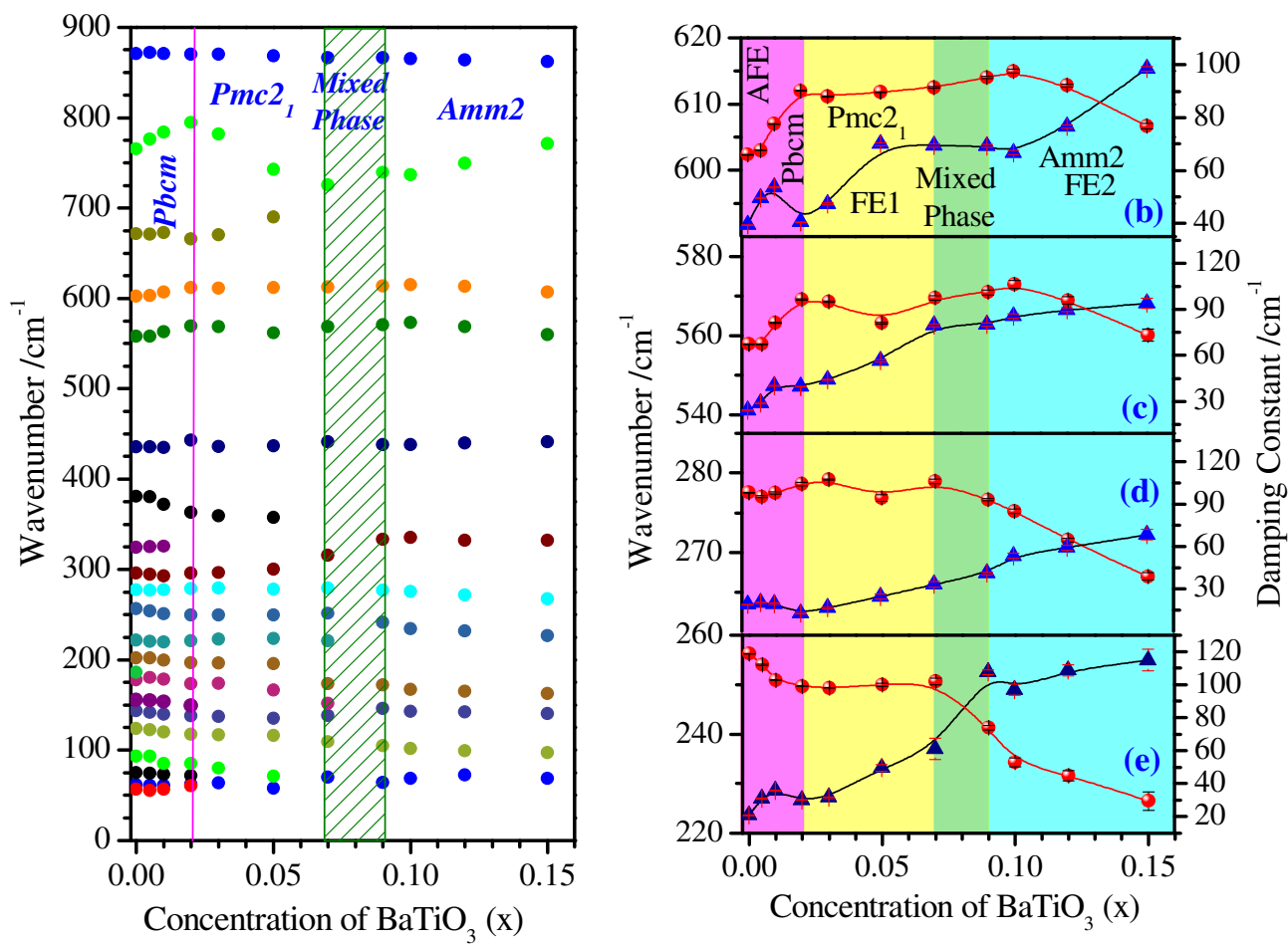

**Figure 3**(a) Raman peak position (wave number/cm$^{-1}$) vs. composition plot for the NNBTx solid solution (x=0.0 ≤x ≤0.15). Dashed lines show the boundary between the structural phases observed during the indexing of x-ray diffraction patterns. Variation of the peak positions (red circles) and damping constant (blue triangles) of stretching modes present at (b) ~601 cm$^{-1}$, (c) ~560 cm$^{-1}$ and bending modes at (d) ~277 cm$^{-1}$, (e) ~255 cm$^{-1}$ are shown in right panel.

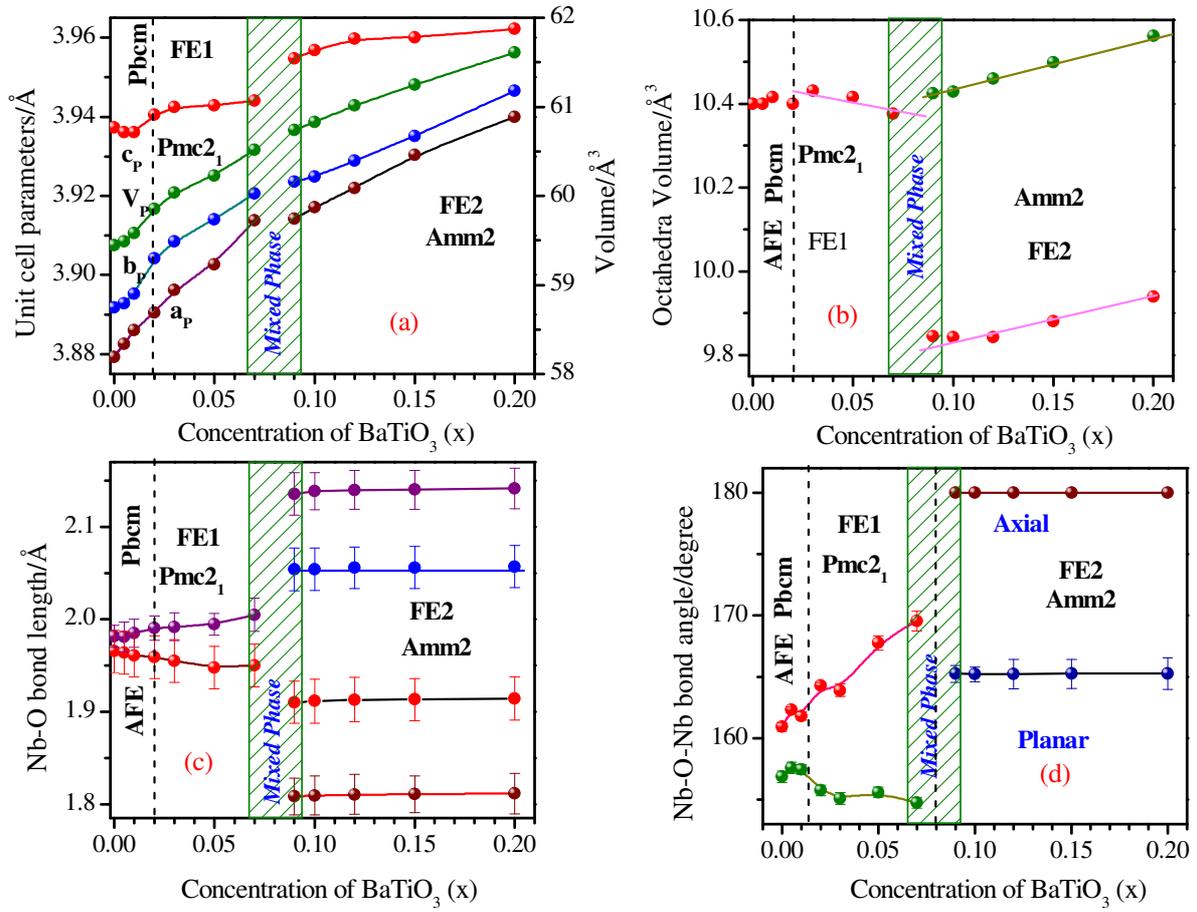

**Figure 4**(a) Variation of cell parameters of different phases in terms of equivalent elementary perovskite cell parameters (ap, bp, and cp) as a function of composition. The relation between elementary perovskite cell and different orthorhombic phases are as: AFE phase ( $A_o= \sqrt{2}\, a_p$, $B_o=\sqrt{2}\, b_p$ and $C_o= 4\, c_p$), FE1 phase ( $A_o= 2\, a_p$, $B_o=\sqrt{2}\, b_p$ and $C_o= \sqrt{2}\, c_p$) and FE2 phase ( $A_o= a_p$, $B_o=2\, b_p$ and $C_o= 2\, c_p$). (b), (c) and (d) show the variation of octahedral volume, axial Nb-O bond lengths and Nb-O-Nb bond angles, respectively.